\documentstyle[12pt,aps,epsf]{revtex}

\input{epsf}
\begin{document}

\title{To the possibility of the "slow light" in the waveguides.}
\author{ Kozlov G.G.}

\maketitle
 \hskip20pt
All Russia Scientific Center "S.I.Vavilov Sate Optical Institute"
 190034, St-Petersburg, Russia
\vskip20pt
\hskip100pt {\it e}-mail:  gkozlov@photonics.phys.spbu.ru
\vskip20pt
\begin{abstract}
The atoms moving within the waveguide with a critical frequency higher than the resonant
frequency of atoms are suggested for obtaining the "slow light".
Due to the absence of the resonant mode in the guide the atoms conserves excitation and coherence.
The speed of this mixed excitation  (electromagnetic field + moving atom) can be very low or even zero.
\end{abstract}

The "slow light" optics is a new trend in physics which appeared during the last decade.
The main problem of the "slow light" physics is to create some medium  in which the speed
of optical pulse can be reduced  up to several tens
meters per second. Medium with a "slow light"  is supposed to be used for storage and
processing of information.
The review of achievements  and questionable points   in "slow light" research  one can find
in \cite{Zap}.
The goal of the present paper is to discuss the  obtaining of
the "slow light"  by means of  atoms moving through  the waveguide with
critical frequency higher than the resonant frequency of atoms.

Let us consider the two-level atom with resonant frequency $\omega_0$ placed into
the waveguide tube whose transverse size $h$ is small enough for critical frequency of
the waveguide $\omega_c$ be higher than the resonant frequency of the atom i.e. $\omega_c>\omega_0$.
Hence the electromagnetic wave with frequency $\omega_0$ is exponentially
 decays (grows) along the waveguide with some characteristic length $\lambda\sim h$ (Fig.1 a).
 Let our two-level atom be excited, so its electrical dipole moment is oscillating at frequency
 $\omega_0$.
Due to the above mentioned property of the waveguide  no radiation decay take place
and the amplitude of oscillations is conserving the same  for infinitely long time.
The oscillating   dipole moment  produce  the electromagnetic field around the
atom but this field goes down to zero  on the right and left hand of the
atom with characteristic length $\lambda$ (Fig.1 b).
So we come to the conclusion that the oscillating dipole within the waveguide with appropriate critical
frequency forms some localised excitation having zero speed and spatial size $\sim\lambda$ along the waveguide.
This excitation is of mixed (atom + electromagnetic field) nature.
The  increasing of  lifetime of the oscillator placed into the
resonator having no resonant mode is wellknown and have been observed experimentally.

If atom has non-zero speed $v$ along the waveguide this  excitation
will travel along the waveguide.
Note that the speed $v$ of this mixed (atom + electromagnetic field)
 excitation can be pretty low or even zero.
 The quantization of this mixed excitation must give the polariton with very low speed of propagation.
 These excitations can be treated as a "slow light" and can be used for creating of
  low speed
optical pulses in the following experiment.
Suppose that  some amount of atoms in the ground state have been prepared near the left end of the waveguide.
Let resonant electromagnetic pulse focused on the left end of the waveguide create
the cloud    of excited atoms near the left end of the guide.
Under the action of this pulse this cloud  acquire  some momentum  and
start moving along the waveguide.
This momentum can be estimated using the balance of photon and atom momentum: $\hbar\omega_0/ c=m v$,
where $c$ - is a speed of ligth, $m$ - mass of the atom, $v$ - speed of the atom.
Because of the small value of photon momentum the speed of atom $v$ can be
 many times smaller  than that of light in vacuum.
Due to the absence of the resonant electromagnetic mode in the waveguide these atoms conserves
optical excitation and coherence and moves slowly to the right end of the guide.
After reaching of the right end of the guide the cloud of excited atoms emits the
electromagnetic pulse similar to that which produced this cloud on the left end
of the guide. The time delay between the initial pulse on the left end of
the guide and the pulse emitted from the right end is determined by the
speed of atoms within the guide and can be very large.

One can control the propagation of excited atoms through the waveguide
 by changing the transverse size of the guide.
 The variation of the transverse size of the waveguide produces the corresponding potential distribution
 for the excited atoms within the guide.
Using this fact one can speed up or slow down the cloud of excited atoms described above.

Note that the waveguide must be empty before the cloud of excited atoms start moving.
It sounds like truth that this cloud of excited atoms within the
 waveguide with $\omega_c>\omega_0$ has much in common with soliton.
What will happens if two optical pulses acts on the left end of the guide in the above experimental setup?
In this case two separated clouds of excited atoms will be injected into the guide. If the
spatial gap between these clouds is larger than $\lambda\sim h$
 the clouds will propagate independently.
So  minimal spatial distance between the clouds  must be $\sim h$.
 The interaction between clouds is a matter of calculation.

If the resonant frequency of the atom in above experiment lays in the optical
region (the corresponding wave length is $\sim 1\mu$m) the transverse size of
the waveguide tube must be less than $1\mu$m.
The losses in the waveguide must be low enough for  dipole moment
 oscillation not to decay during atoms travels from one end of the guide to another.
It seems impossible to create such a device at present time  but
 the achievements of modern technology (the photonic crystals technology)
 gives some hopes that it will become possible in nearest future.

Be free to comment and criticize all the above statements in online regime.

\begin{figure}
\epsfxsize=400pt
\epsffile{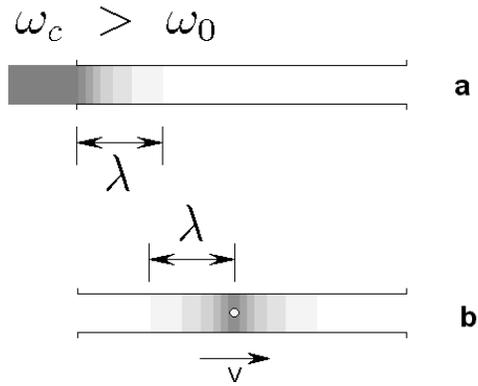}
\caption{ a. The electromagnetic field at frequency $\omega_0$
 goes down to zero within the guide whose critical frequency
$\omega_c>\omega_0$, b. The atom
 whose dipole moment is oscillating at frequency $\omega_0<\omega_c$
conserve its excitation and produce the localized electromagnetic field within the guide.
This excited atom can travel along the guide and introduce the "slow light".}
\end{figure}
\end{document}